# Manifestation of Extremely High-Q Pseudo-Modes in Scattering of a Bessel Light Beam by a Sphere

Vasily Klimov[1]

[1] *P.N. Lebedev Physical Institute, Russian Academy of Sciences,*

*53 Leninsky Prospekt, Moscow 119991, Russia*

e-mail address: klimov256@gmail.com

*The exact analytical solution of Maxwell equations for a Bessel light beam scattered by a sphere is found. Scattered power, stored energy and a generalized Q factor as a function of frequency, the sphere radius, permittivity and the Bessel beam angle are found. On the base of this solution, modes and pseudo-modes of a dielectric sphere are extracted by calculation of the generalized Q factor. It is shown that an appropriate choice of Bessel beam parameters can provide excitation of a single given mode and an unlimited value of the radiative Q factor of pseudo-modes.*

      For many optical devices, it becomes critical to localize electromagnetic fields in subwavelength volumes.  In particular, it is extremely important for modern applications in nanophotonics including lasing and spasing [1], nanoantennas [2], metasurfaces [3,4,5], sensing [6, 7], optical computing [8,9], and applications within quantum optics and topological photonics [10-12].

      Due to radiative losses, the physics and the description of high-quality resonant modes in nanoresonators are rather complicated and as far as we know the only known analytical solution is the one for Mie scattering of a plane

electromagnetic wave by a sphere [13,14]. Within this solution, the scattering cross-section looks like

$$\sigma_{sca} = \frac{2\pi}{k_0^2} \sum_{n=1}^{\infty} (2n+1)\left(|a_n|^2 + |b_n|^2\right), \qquad (1)$$

where $a_n$ and $b_n$ are Mie reflection coefficients [14,15] (see also (9) and (17)). The analytical solution (1) is very important and has been applied for innumerous studies. In particular, recently this solution has been applied to search optimum forward light scattering [16], anomalous light scattering [17], non-radiating anapole states [18,19], and nanoparticles with pure high-order multipoles [20].

However, in the solution (1) contributions from different TE and TM modes cannot be separated, and arbitrary small scattering ("invisibility") and pure high-order multipole nanoparticles cannot be fully realized. This is due to the fact that when Mie scattering coefficient for the specific mode equals to zero, other Mie coefficients are not zero, and the total scattering still remains nonzero.

However, now many other types of light beams are of interest for optical community (see e.g. [ 21-27]). Non-diffracting Bessel beams are among them [28, 29]. Scattering of acoustic Bessel beams by a sphere is presented in [30].

But as far as we know no exact solution for scattering of Bessel beams by a sphere is known for electromagnetic case, and the goal of this work is to fill this gap. In this work, we have found an exact analytical solution for Bessel light beam scattering on a sphere for the first time and an analytical expression for the generalized Q factor,

$$Q = \frac{\omega W_{st}}{P_{sc}} \qquad (2)$$

where $W_{st}$ and $P_{sc}$ are the stored energy and the scattered power, correspondingly. The generalized $Q$ factor (2) is defined for any parameters of the problem, and for eigen frequencies it gives the value of $Q$ factor of eigen-modes. It is of additional interest in the generalized $Q$ factor that it allows one to find light states, which are not eigen-modes, but have extra high $Q$ factors.

More specifically, we have solved Maxwell equations for a sphere excited by an axisymmetric Bessel beam of zero order

$$H_\varphi = H_0 J_1(k_0 \sin\beta \rho) e^{ik_0 z \cos\beta} \text{ (TM case)}, \qquad (3)$$

$$E_\varphi = E_0 J_1(k_0 \sin\beta \rho) e^{ik_0 z \cos\beta} \text{ (TE case)}. \qquad (4)$$

The wave vector components of the Bessel beams (3) and (4) form a cone having a conical angle β relative to the $z$ axis. Our solution is valid for arbitrary parameters, but here for simplicity we will consider a lossless dielectric sphere of a radius $a$ in vacuum, where $k_0 = \omega/c$. The geometry of the problem is shown in Fig.1.

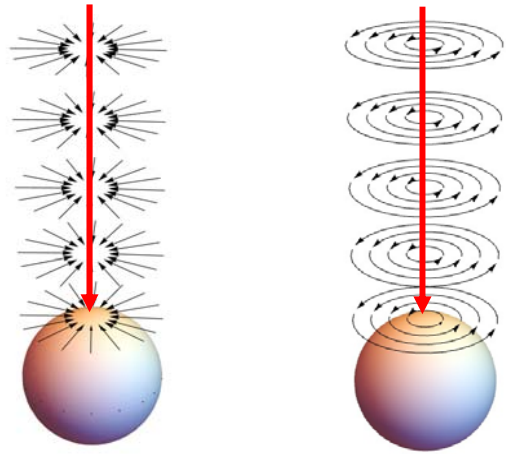

Fig. 1. The geometry of the problem: a) TM case, b) TE case. The sphere radius is equal $a$.

In spherical coordinates $z = R\cos\theta, \rho = R\sin\theta$, we have instead of (3):

$$H_\varphi(R,\theta) = H_0 J_1(k_0 R \sin\theta \sin\beta) e^{ik_0 R \cos\theta \cos\beta} \qquad (5)$$

Using symmetry (5) with respect to permutations of $\theta$ and $\beta$, it can be shown that the expansion of (5) in spherical harmonics has the form

$$H_\varphi(R,\theta) = -iH_0 \sum_{n=1}^{\infty} i^n \frac{2n+1}{n(n+1)} j_n(k_0 R) \frac{\partial P_n(\cos\theta)}{\partial \theta} \frac{\partial P_n(\cos\beta)}{\partial \beta} \qquad (6)$$

In (6), $j_n(z)$ and $P_n(\cos\theta)$ are the spherical Bessel functions and the Legendre polynomial, correspondigly. Expansion (6) is crucial for this work.

Using (6), the expressions for fields can be written as

$$H_\phi^R(R,\theta) = -iH_0 \sum_{n=1}^{\infty} a_n i^n \frac{2n+1}{n(n+1)} h_n(k_0 R) \frac{\partial P_n(\cos\theta)}{\partial \theta} \frac{\partial P_n(\cos\beta)}{\partial \beta}$$

$$E_\theta^R(R,\theta) = -\frac{H_0}{k_0 R} \sum_{n=1}^{\infty} a_n i^n \frac{2n+1}{n(n+1)} \left[z h_n(z)\right]'_{z=k_0 R} \frac{\partial P_n(\cos\theta)}{\partial \theta} \frac{\partial P_n(\cos\beta)}{\partial \beta} \quad (7)$$

for scattered fields and as

$$H_\phi^T(R,\theta) = -iH_0 \sum_{n=1}^{\infty} d_n i^n \frac{2n+1}{n(n+1)} j_n(k_1 R) \frac{\partial P_n(\cos\theta)}{\partial \theta} \frac{\partial P_n(\cos\beta)}{\partial \beta}$$

$$E_r^T(R,\theta) = -\frac{H_0}{k_0 \varepsilon_1 R} \sum_{n=1}^{\infty} d_n i^n (2n+1) j_n(k_1 R) P_n(\cos\theta) \frac{\partial P_n(\cos\beta)}{\partial \beta} \quad (8)$$

$$E_\theta^T(R,\theta) = -\frac{H_0}{k_0 \varepsilon_1 R} \sum_{n=1}^{\infty} d_n i^n \frac{2n+1}{n(n+1)} \left[z j_n(z)\right]'_{z=k_1 R} \frac{\partial P_n(\cos\theta)}{\partial \theta} \frac{\partial P_n(\cos\beta)}{\partial \beta}$$

for fields inside the sphere. In (8) and further, $k_1 = \sqrt{\varepsilon} k_0$ stands for the wave number inside the sphere.

The continuity condition for the tangential components of the fields allows us to find the coefficients $a_n$ and $d_n$

$$a_n = -\frac{\varepsilon_1 j_n(z_1)\left[z_0 j_n(z_0)\right]' - j_n(z_0)\left[z_1 j_n(z_1)\right]'}{\varepsilon_1 j_n(z_1)\left[z_0 h_n(z_0)\right]' - h_n(z_0)\left[z_1 j_n(z_1)\right]'} =$$

$$d_n = -\varepsilon_1 \frac{h_n(z_0)\left[z_0 j_n(z_0)\right]' - j_n(z_0)\left[z_0 h_n(z_0)\right]'}{\varepsilon_1 j_n(z_1)\left[z_0 h_n(z_0)\right]' - h_n(z_0)\left[z_1 j_n(z_1)\right]'} \quad (9)$$

where $z_{0,1} = k_{0,1} a$. Naturally, the calculation results correspond to the Mie coefficient for the TM case.

Knowing the fields outside and inside the sphere, one can find the scattered power:

$$P_{sc} = \frac{cH_0^2}{2k_0 k_2} \sum_{n=1}^{\infty} |a_n|^2 \frac{2n+1}{n(n+1)} \left( \frac{\partial P_n(\cos\beta)}{\partial \beta} \right)^2 \qquad (10)$$

and the energy stored in the sphere:

$$W_{st} = \frac{1}{16\pi} \int_V dV \left[ |H_\varphi^T|^2 + \varepsilon |E_\theta^T|^2 + \varepsilon |E_r^T|^2 \right] = \\ \frac{H_0^2}{4k_1^3} \sum_{n=1}^{\infty} |d_n|^2 \frac{2n+1}{n(n+1)} \left( \frac{\partial P_n(\cos\beta)}{\partial \beta} \right)^2 I(n, z_1) \qquad (11)$$

where

$$I(n,z) = z\left(1 + n + z^2\right) j_n(z)^2 - z^2 \left(z_1 j_{n-1}(z) + j_n(z)\right) j_{n+1}(z) \qquad (12)$$

An important feature of any resonators is their Q factor of eigen- oscillation, which is determined by the imaginary part of eigen-frequency which in its turn is a root of the denominator of the Mie coefficients (9).

However, in scattering problems, the appearance of new interesting states with low radiation losses that are not directly related to eigen-modes, is possible. To find such states, it is natural to use the concept of a generalized Q factor [31] which is defined for any parameters of the problem, and at each resonant frequency $\omega_n$ it attains the value of the corresponding $Q$ factor.

In the TM case, for a generalized $Q$ factor, we have the following expression:

$$Q^{TM} = \frac{\omega W_{st}}{P_{sc}} = \frac{1}{2\varepsilon^{3/2}} \frac{\sum_{n=1}^{\infty} |d_n|^2 \frac{2n+1}{n(n+1)} \left( \frac{\partial P_n(\cos\beta)}{\partial \beta} \right)^2 I(n, z_1)}{\sum_{n=1}^{\infty} |a_n|^2 \frac{2n+1}{n(n+1)} \left( \frac{\partial P_n(\cos\beta)}{\partial \beta} \right)^2} \qquad (13)$$

In TE case, the calculations are completely similar, and for the scattered power, the stored energy, and the generalized $Q$ factor, we have the following expressions:

$$P_{sc} = \frac{cE_0^2}{2k_0^2} \sum_{n=1}^{\infty} |b_n|^2 \frac{2n+1}{n(n+1)} \left( \frac{\partial P_n(\cos\beta)}{\partial \beta} \right)^2 \qquad (14)$$

$$W_{st} = \frac{E_0^2}{4k_0^2 k_1} \sum_{n=1}^{\infty} |c_n|^2 \frac{2n+1}{n(n+1)} \left( \frac{\partial P_n(\cos\beta)}{\partial \beta} \right)^2 I(n, z_1) \qquad (15)$$

$$Q^{TE} = \frac{1}{2\sqrt{\varepsilon}} \cdot \frac{\sum_{n=1}^{\infty} |c_n|^2 \frac{2n+1}{n(n+1)} \left( \frac{\partial P_n(\cos\beta)}{\partial \beta} \right)^2 I(n, z_1)}{\sum_{n=1}^{\infty} |b_n|^2 \frac{2n+1}{n(n+1)} \left( \frac{\partial P_n(\cos\beta)}{\partial \beta} \right)^2} \qquad (16)$$

In (14), (15), and (16), $b_n$ and $c_n$ are Mie coefficients for TE modes

$$b_n = -\frac{j_n(z_1)[z_2 j_n(z_2)]' - j_n(z_0)[z_1 j_n(z_1)]'}{j_n(z_1)[z_0 h_n(z_0)]' - h_n(z_0)[z_1 j_n(z_1)]'}$$

$$c_n = -\frac{h_n(z_0)[z_0 j_n(z_0)]' - j_n(z_0)[z_0 h_n(z_0)]'}{j_n(z_1)[z_0 h_n(z_0)]' - h_n(z_0)[z_1 j_n(z_1)]'} \qquad (17)$$

The solutions found are very remarkable, because, in contrast to the classical solution (1), they contain an additional parameter $\beta$ (a conical angle of the Bessel beam) by changing which one can control the interaction of the light beam with specific modes in wide limits. In particular, if $\beta$ is chosen so that $\frac{\partial P_{\bar{n}}(\cos\beta)}{\partial \beta} = 0$ the mode with the number $\bar{n}$ will not be excited at all, and the total scattering will decrease correspondingly.

As an example of the application of the solutions found, the dependences on the size parameter $k_0 a$ of the scattered power, the stored energy, and the generalized $Q$ factor for TM polarization and $\varepsilon = 36$ (PbTe [27]) are shown in Fig.2. For TE polarization, the results are similar.

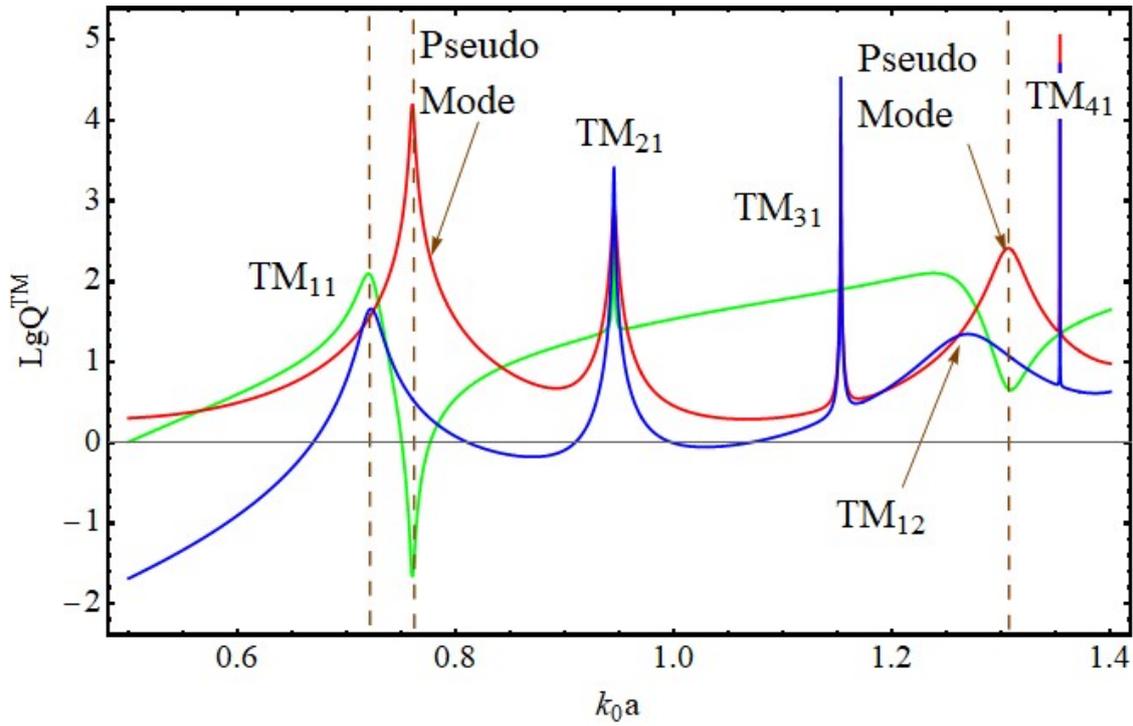

Fig. 2. The dependence of the scattered power (green curve), the stored energy (blue curve), and the generalized $Q$ factor (red curve) on the size parameter $k_0 a$ (TM case, $\varepsilon = 36$, $\beta = \pi / 4$).

An analysis of Fig. 2 shows that in the spectrum of the generalized $Q$ factor (red curve) there are well-defined maxima which positions do not always coincide with the positions of the eigen-modes. We referred to these configurations as pseudo-modes since there are no real natural oscillation for these parameters. In this case, the maximums of generalized $Q$ factor corresponding to pseudo-modes are related to a decrease in the scattered power (green curve). The position of true eigen-modes can be found from the spectrum of the energy stored inside the sphere (blue curve).

The solution found allows one to find conditions under which pseudo-modes will have radiative losses substantially smaller and generalized $Q$ factors substantially greater than true eigen-modes.

Even smaller radiation and a higher generalized $Q$ factor of the pseudo-modes can be obtained by considering the superposition of Bessel beams:

$$H_\varphi(R,\theta) = H_0 \sum_i \alpha_i J_1(k_0 R \sin\theta \sin\beta_i) e^{ik_0 R \cos\theta \cos\beta_i}$$

$$= -iH_0 \sum_{n=1}^{\infty} i^n \frac{2n+1}{n(n+1)} j_n(k_0 R) \frac{\partial P_n(\cos\theta)}{\partial\theta} \Psi_n \quad (18)$$

$$\Psi_n = \sum_i \alpha_i \frac{\partial P_n(\cos\beta_i)}{\partial\beta_i}$$

A special choice of the coefficients $\alpha_1, \alpha_2, \alpha_3 \ldots$ and the angles $\beta_1, \beta_2, \beta_3 \ldots$ of Bessel beams allows in principle one to excite the only ONE specific mode and obtain unlimited radiation $Q$ factors, since for any given mode the Mie coefficient is equal to zero for some real frequency values.

In Figure 3, the generalized $Q$ factor and the scattered power for a sphere excited by single Bessel beams with $\beta = \pi/4, 3\pi/8, 4\pi/9, \pi/2$ and by superposition of 4 Bessel beams (see (18)) with $\alpha = -5.25, 7.43, -5.25, 2.61$ and $\beta = \pi/8, \pi/4, 3\pi/8, \pi/2$ are shown as a function on the size parameter $k_0 a$ at $\varepsilon = 36$. Such a choice of superposition parameters allows one to suppress excitation of multipoles with $n=2,3,4$ ($\Psi_n=0$ for $n=2,3,4$). Thus, in the region of zero scattering of dipole $TM_{21}$ modes, the total radiation will be determined by modes with multipole orders $n \geq 5$ (dotriacontapole, etc.) and therefore will be extremely small.

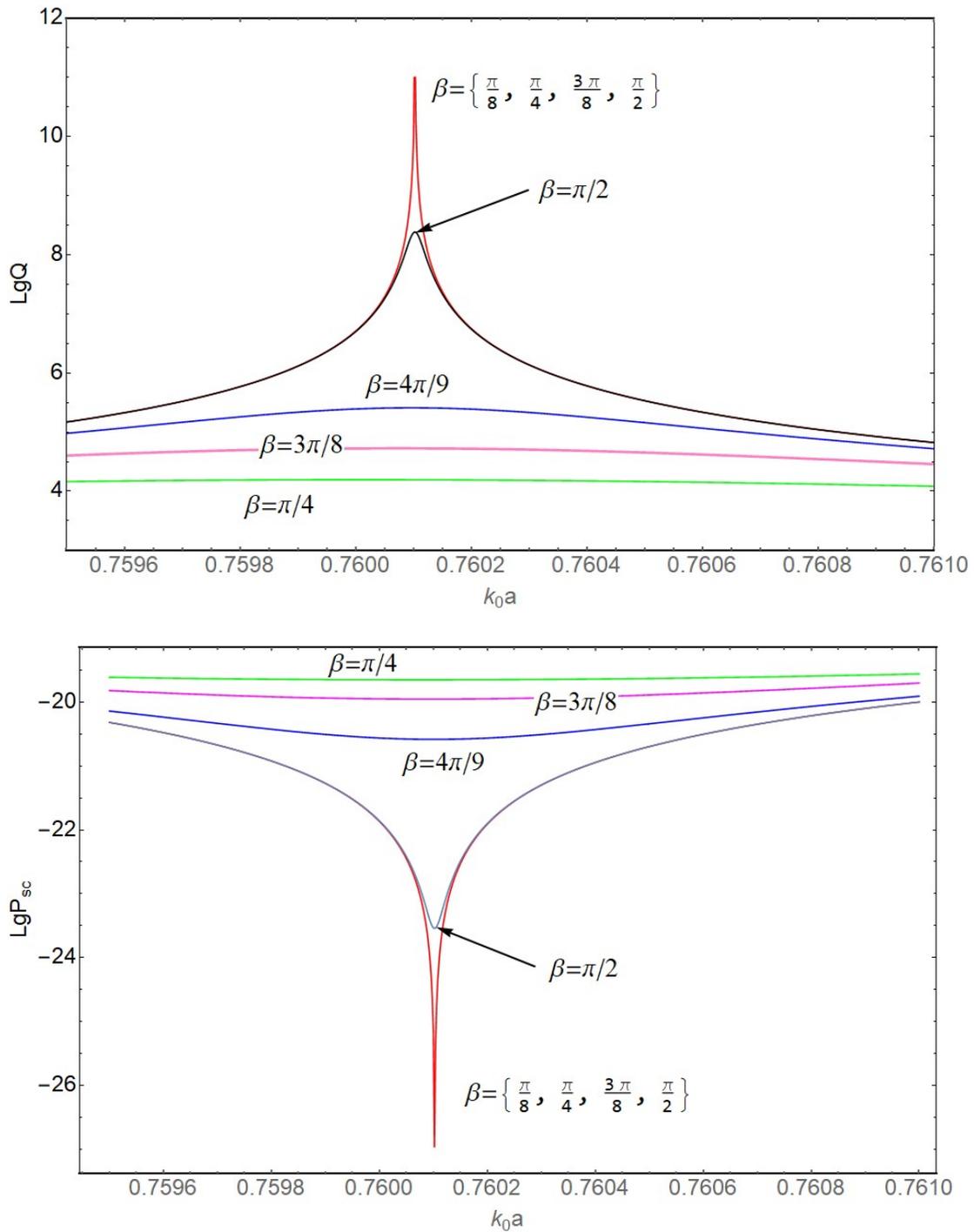

Fig. 3. Dependence of the generalized $Q$ factor (a) and the scattered power (b) for a sphere excited by single Bessel beams with $\beta = \pi/4, 3\pi/8, 4\pi/9, \pi/2$ and by a superposition of 4 Bessel beams (see(18)) with $\alpha = -5.25, 7.43, -5.25, 2.61$ and $\beta = \pi/8, \pi/4, 3\pi/8, \pi/2$ on the size parameter (TM case, $\varepsilon = 36$).

It can be seen from Fig. 3 that, as a result of optimizing the parameters of Bessel beams, the generalized $Q$ factor can be enhanced by more than 6 orders of magnitude in comparison with the general case. Such an effect cannot be achieved using excitation by a plane wave (1).

Figure 4 shows the distribution of the magnetic field strength and streamline of the Poynting vector when the sphere is excited by a Bessel beam with $\beta=4\pi/9$(a) and a superposition of 4 Bessel beams with $\beta = (\pi/8, \pi/4, 3\pi/8, \pi/2)$ ($k_0a = 1.31$, $\varepsilon = 36$).

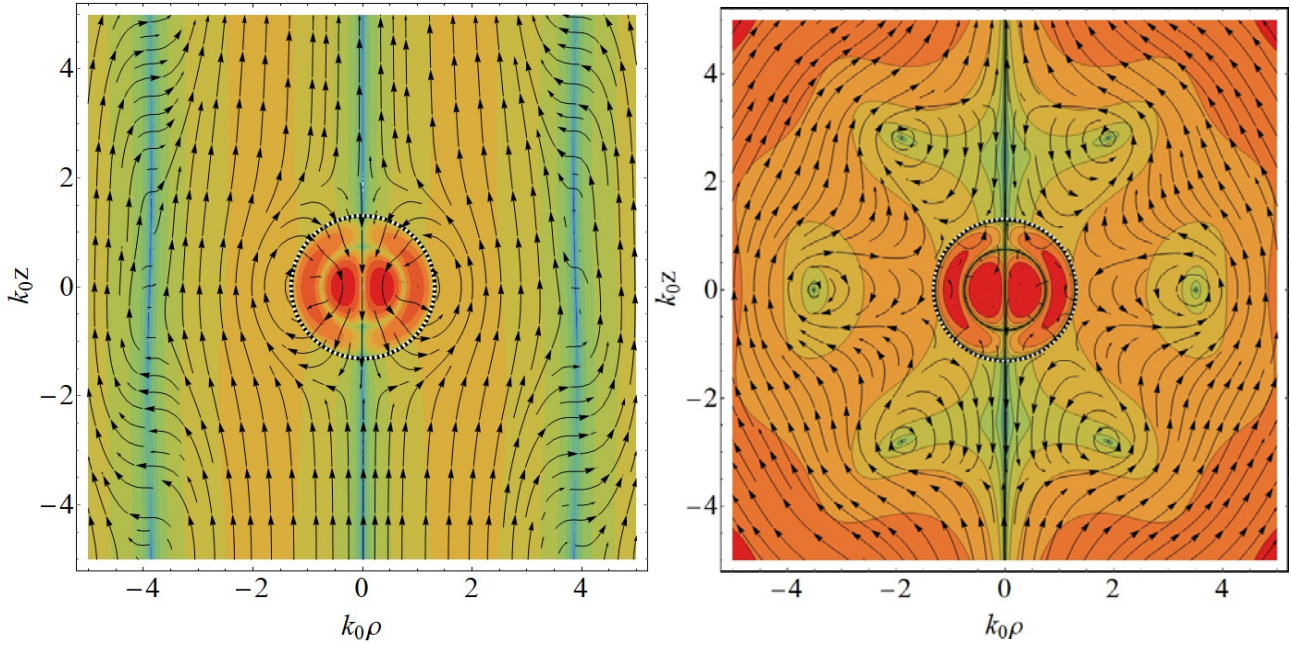

Fig. 4. Distribution of the magnetic field strength ($ln\ |H_\varphi|$) and the streamline of the Poynting vector when the sphere is excited by a Bessel beam with $\beta=4\pi/9$(a) and a superposition of 4 Bessel beams $\beta = (\pi/8, \pi/4, 3\pi/8, \pi/2)$ (b). ($k_0a =1.31$, $\varepsilon=36$).

Fig.4 shows that, indeed, the field distribution outside the sphere upon excitation of the sphere by superposition of 4 Bessel beams has a significantly higher multipolar order $n \geq 5$ (dotriacontapole, etc.), compared with the dipole character of the field inside the sphere ($n = 1$ for the $TM_{12}$ mode). This circumstance leads to a radical enhancement of the $Q$ factor of pseudo-modes.

A further increase in the number of interfering Bessel beams will allow one to increase the $Q$ factor unlimitedly, since in fact, only one term will remain in series (18), which corresponds to the excitation of the sphere by a spherical converging wave, for which the complete absence of scattering is possible when the corresponding Mie scattering coefficient vanishes.

Note that a similar unlimited increase in the quality factor is found in the case of bound states in continuum [33, 34]. However, the physics of this phenomena is completely different from our case, since in the case of bound states in continuum [33, 34] the true eigen-oscillations of a two-dimensional system exist, while in our case we are talking about pseudo-modes that are absent in the absence of external fields.

**Conclusion**

Thus, in the present work, an exact solution of the Maxwell equation, which describes the excitation of a sphere by a Bessel beam, was found. Based on this solution, a generalized $Q$ factor was found, and it was shown that together with the usual eigen-modes, the spectrum of a generalized $Q$ factor contains pseudo-modes that are associated with a decrease in the scattered power. It is shown that with a special choice of the parameters of Bessel beams, it is possible to excite only one given mode or pseudo-mode and an almost unlimited decrease in the dissipated power and, as a result, an unlimited increase in the generalized $Q$ factor of pseudo-modes can be achieved.

In this work, we have considered a homogeneous sphere in the superposition field of Bessel beams, however, all the results obtained can be directly generalized to the case of spherically layered structures.

Since using the developed approach it is possible to excite given modes or pseudo-modes of arbitrary multipolarity with very narrow resonance widths and extremely high fields inside a sphere, the results obtained pave the way for creating new optical systems and devices. In particular, we believe that our results lay foundation for development of nanolasers, biosensors, nonlinear and quantum optical chips.

Acknowledgment

This work was supported by the Russian Foundation for Basic Research (Grant No. 18-02-00315).